\documentclass[onecolumn,prd,amsmath,amssymb,nofootinbib]{revtex4}
\usepackage{graphicx}
\usepackage{dcolumn}
\usepackage{bm}
\usepackage{hyperref}

\begin{document}

\title{Comment on ``Application of the three-dimensional telegraph 
equation to cosmic-ray transport'' (arXiv:1606.08272)}

\author{Andrei Galiautdinov}
\email{ag1@uga.edu}
\affiliation{
Department of Physics and Astronomy, 
University of Georgia, Athens, Georgia 30602, USA
}

\date{\today}

\begin{abstract}
In a recent publication [R.\ C.\ Tautz and I.\ Lerche, Res.\ Astron.\ Astrophys.\ 
{\bf 16}, 162 (2016); arXiv:1606.08272], the authors present a derivation 
of the Green's function for the three-dimensional telegraph equation
(also known as the heat wave equation, or relativistic heat conduction 
equation). We demonstrate that the closed-form expression derived in 
their Appendix A is incorrect. Specifically, the published solution lacks 
a Dirac delta term representing the ballistic wavefront and contains an 
algebraic error in the prefactor of the wake term. These omissions arise 
from the neglect of distributional derivatives when differentiating a 
Heaviside step function. We provide a rigorous derivation of the Green's 
function using the Fourier transform method, verify the correct limiting 
behavior as the damping vanishes, and pinpoint the exact mathematical 
step where the original derivation failed.
\end{abstract}

\maketitle

\section{Introduction}

The three-dimensional telegraph equation (also known as the heat 
wave equation, or relativistic heat conduction equation),
\begin{equation}
\label{eq:TEq}
\left(
\frac{\partial^2}{\partial{t}^2} 
+2{\beta}\frac{\partial }{\partial{t}}
-\nabla^2
\right)f(\vec{r},t)
=  S(\vec{r},t)\, ,
\end{equation}
where positive ${\beta}$ plays the role of the ``damping'' coefficient, 
describes transport phenomena that exhibit both diffusive and 
wave-like characteristics, ensuring finite propagation speeds. 
Tautz and Lerche \cite{Tautz2016} applied this equation to cosmic-ray 
transport, deriving an analytical solution for the Green's function in 
their Appendix A. However, the resulting Green's function presented 
in Eq.~(A.25) of Ref.~\cite{Tautz2016} is physically inconsistent: 
it vanishes or becomes smooth in the limit where damping goes 
to zero, failing to recover the causal Green's function of the standard 
wave equation (which requires a Dirac delta singularity).

In this Comment, we present a complete, corrected derivation of 
the Green's function and explicitly identify the calculus error in the 
original paper.

\section{Derivation of the Green's Function}
\label{sec:Derivation}

\subsection{General solution}
\label{sec:GF}

The general solution of Eq.\ (\ref{eq:TEq}) can be found by 
the Fourier transform method. We first write,
\begin{align}
f(\vec{r},t)
=
\int \frac{d^3 \vec{k}}{(2\pi)^3}
e^{i\vec{k}\cdot \vec{r}} f(\vec{k},t),
\quad 
S(\vec{r},t)
=
\int \frac{d^3 \vec{k}}{(2\pi)^3}
e^{i\vec{k}\cdot \vec{r}}S(\vec{k},t),
\end{align}
which upon substitution into (\ref{eq:TEq}) gives
\begin{equation}
\label{eq:TEqk}
\left(
\frac{\partial^2}{\partial{t}^2} 
+2{\beta}\frac{\partial }{\partial{t}}
+ |\vec{k}|^2
\right)f(\vec{k},t)
= S(\vec{k},t).
\end{equation}
We next write,
\begin{align}
f(\vec{k},t)
=
\int \frac{d^3 \omega}{2\pi}
e^{-i\omega t} f(\vec{k},\omega),
\quad 
S(\vec{k},t)
=
\int \frac{d^3 \omega}{2\pi}
e^{-i\omega t}S(\vec{k},\omega),
\end{align}
and after plugging into (\ref{eq:TEqk}) get,
\begin{equation}
\label{eq:TEqkw}
\left(
\omega^2 
+2i{\beta}\omega
- |\vec{k}|^2
\right)f(\vec{k},\omega)
= -S(\vec{k},\omega),
\end{equation}
whose general (in the distributional sense) solution is given by
\begin{align}
f(\vec{k},\omega)
=
-\frac{S(\vec{k},\omega)}
{(\omega-\omega^{(1)}_{\vec{k}})(\omega-\omega^{(2)}_{\vec{k}})}
+
(2\pi)^4 C_1(\vec{k})\delta(\omega -\omega^{(1)}_{\vec{k}})
+
(2\pi)^4 C_2(\vec{k})\delta(\omega -\omega^{(2)}_{\vec{k}}),
\end{align}
where the free-mode frequencies are
\begin{equation}
\omega^{(1,2)}_{\vec{k}}=-i{\beta}\pm \sqrt{|\vec{k}|^2-{\beta}^2}.
\end{equation}

Thus, the general solution of our three-dimensional telegraph equation
with \emph{constant} $\beta$ is
\begin{align}
f(\vec{r},t)
&=
\int_{-\infty}^{\infty} dt'
\int d^{3} \vec{r'} G(\vec{r},t;\vec{r'},t')S(\vec{r'},t')
\nonumber \\
& \qquad
+
\int \frac{d^3 \vec{k}}{(2\pi)^3}
\left[
C_1(\vec{k}) e^{-i(\omega^{(1)}_{\vec{k}}t-\vec{k}\cdot \vec{r})}
+
C_2(\vec{k}) e^{-i(\omega^{(2)}_{\vec{k}}t-\vec{k}\cdot \vec{r})}
\right],
\end{align}
where the Green's function is given by
\begin{equation}
\label{eq:Grtr't'}
G(\vec{r},t;\vec{r'},t')
=
\frac{1}{(2\pi)^3}
\int d^3 \vec{k} e^{-i\vec{k}\cdot \vec{r}} G(t,t',\vec{k}),
\end{equation}
with
\begin{equation}
\label{eq:Gtt'k}
G(t,t',\vec{k})
=
-\frac{1}{2\pi}
\int_{-\infty}^{\infty} d \omega 
\frac{e^{-i \omega (t-t')}}{(\omega-\omega^{(1)}_{\vec{k}})(\omega-\omega^{(2)}_{\vec{k}})}.
\end{equation}

Notice that with positive ${\beta} >0$ we have four types of 
poles in the Green's  function, all of which (except for $\omega=0$) 
reside below the real axis in the complex $\omega$-plane:
\begin{itemize}
\item{Type 1:} $|\vec{k}|^2>{\beta}^2$, giving 
$\omega^{(1,2)}=-i{\beta}\pm \sqrt{|\vec{k}|^2-{\beta}^2}$,
with ${\rm Im}(\omega^{(1,2)})=-{\beta} <0$;
\item{Type 2:} $|\vec{k}|^2={\beta}^2$, giving 
$\omega^{(1,2)}=-i{\beta}$,
with ${\rm Im}(\omega^{(1,2)})=-{\beta} <0$;
\item{Type 3:} $|\vec{k}|^2<{\beta}^2$, giving 
$\omega^{(1,2)}=-i{\beta}\pm i\sqrt{{\beta}^2-|\vec{k}|^2}$,
with ${\rm Im}(\omega^{(1,2)})=-{\beta} 
\pm \sqrt{{\beta}^2-|\vec{k}|^2}<0$;
\item{Type 4:} $|\vec{k}|^2=0$, giving 
$\omega^{(1)}=0$, with ${\rm Im}(\omega^{(1)})=0$, 
and $\omega^{(2)}=-2i{\beta}$,
with ${\rm Im}(\omega^{(2)})=-2{\beta}<0$.
\end{itemize}

Performing the contour integration in Eq.\ (\ref{eq:Gtt'k}), where 
only the vanishing pole (of Type 4) needs to be shifted below 
the real axis in the complex $\omega$-plane (which, essentially, 
means that in our theory the Green's function is automatically 
retarded for $\beta>0$, corresponding to a future-directed temporal 
orientation on the spacetime manifold, in contrast to usual 
electrodynamics, where such choice must be made by hand) we first 
find,
\begin{equation}
\label{eq:Gtt'kFOUND}
G(t,t',\vec{k})
=
\theta(t-t')\frac{1}{i}
\left(
\frac{e^{-i\omega^{(2)}_{\vec{k}}(t-t')}}{\omega^{(1)}_{\vec{k}}-\omega^{(2)}_{\vec{k}}}
-
\frac{e^{-i\omega^{(1)}_{\vec{k}}(t-t')}}{\omega^{(1)}_{\vec{k}}-\omega^{(2)}_{\vec{k}}}
\right)
=
\frac{\theta(t-t')e^{-{\beta}(t-t')}
\sin\left(\sqrt{k^2-{\beta}^2} \, (t-t')\right)}
{\sqrt{k^2-{\beta}^2}},
\end{equation}
where $\theta(t-t')$ is the Heaviside step function, and $k\equiv |\vec{k}|$. 
With this, Eq.\ (\ref{eq:Grtr't'}) becomes 
\begin{align}
G(\vec{r},t;\vec{r'},t') 
&=
\frac{\theta(T)e^{-{\beta}T}}{(2\pi)^3}
\int d^3 \vec{k} \;
e^{i\vec{k} \cdot \vec{R}} \;
\frac{\sin\left(\sqrt{k^2-{\beta}^2} \, T\right)}
{\sqrt{k^2-{\beta}^2}},
\end{align} 
where $T\equiv t-t'$, $\vec{R}\equiv \vec{r}-\vec{r'}$ and $R\equiv |\vec{R}|$.
Integration over the angles gives
\begin{equation}
\label{eq:Gintegral1}
G(R,T)
=
\frac{ \theta(T)e^{-{\beta}T}}{2\pi^2 R}
\int_0^{+\infty} dk \;
\frac{k \sin\left(\sqrt{k^2-{\beta}^2} \, T\right)\sin(kR)}
{\sqrt{k^2-{\beta}^2}}.
\end{equation}

To evaluate the integral,
\begin{equation}
I(R,T;\beta) = \int_0^\infty k \, \sin(kR) \, 
\frac{\sin\big(T \sqrt{k^2-\beta^2}\big)}{\sqrt{k^2-\beta^2}} \, dk,
\end{equation}
we first note the identity
\begin{equation}
k \, \sin(kR) = - \frac{\partial}{\partial R} \cos(kR),
\end{equation}
so that the integral becomes
\begin{equation}
I(R,T;\beta) = - \frac{\partial}{\partial R} \int_0^\infty \cos(kR) \, 
\frac{\sin\big(T \sqrt{k^2-\beta^2}\big)}{\sqrt{k^2-\beta^2}} \, dk.
\end{equation}
The closely related integral,
\begin{equation}
F(R,T) = \int_0^\infty \cos(kR) \, \frac{\sin\big(T \sqrt{k^2 + m^2}\big)}{\sqrt{k^2 + m^2}} \, dk ,
\end{equation}
is known from standard tables (e.g., Gradshteyn \& Ryzhik, Formula 3.876.1),
\begin{equation}
F(R,T) = \frac{\pi}{2} J_0\big(m \sqrt{T^2 - R^2}\big) \, \theta(T-R),
\end{equation}
whose step-by-step proof may be found here: 
\url{https://math.stackexchange.com/questions/1376303/derivation-of-gradshteyn-and-ryzhik-integral-3-876-1-in-question}.

In our case, the integral has $\sqrt{k^2-\beta^2}$ instead of 
$\sqrt{k^2 + m^2}$. The two expressions are related via the 
substitution $m = i \beta$. Using the identity $J_0(iz) = I_0(z)$, 
where $I_0$ is the modified Bessel function of the first kind, we 
get,
\begin{equation}
\int_0^\infty \cos(kR) \, \frac{\sin\big(T \sqrt{k^2-\beta^2}\big)}{\sqrt{k^2-\beta^2}} \, dk
= \frac{\pi}{2} I_0\big(\beta \sqrt{T^2 - R^2}\big) \, \theta(t-r).
\end{equation}
Applying the derivative $- \frac{\partial}{\partial R}$, we obtain
\begin{equation}
I(R,T;\beta) = - \frac{\pi}{2} \frac{\partial}{\partial R} \Big[ I_0\big(\beta \sqrt{T^2 - R^2}\big) \, \theta(T-R) \Big].
\end{equation}
Using the product rule and identities
\begin{equation}
\frac{d}{dR} \theta(T-R) = - \delta(T-R), \qquad
\frac{d}{dz} I_0(z) = I_1(z),
\end{equation}
we have
\begin{align}
\frac{\partial}{\partial R} \Big[ I_0\big(\beta \sqrt{T^2 - R^2}\big) \, \theta(T-R) \Big]
&= I_1\big(\beta \sqrt{T^2 - R^2}\big) \frac{\partial}{\partial R} 
\big(\beta \sqrt{T^2 - R^2}\big) \theta(T-R) + I_0(0) \, \frac{d}{dR} \theta(T-R) \\
&= I_1\big(\beta \sqrt{T^2 - R^2}\big) 
\left( - \frac{\beta R}{\sqrt{T^2 - R^2}} \right) \theta(T-R) - \delta(T-R),
\end{align}
since $I_0(0) = 1$. Including the prefactor $- \pi/2$, we arrive at 
\begin{equation}
\label{eq:Irtbeta}
I(R,T;\beta) = \frac{\pi}{2} \, 
\delta(T-R) + \frac{\pi \beta R}{2 \sqrt{T^2 - R^2}} \, I_1\big(\beta \sqrt{T^2 - R^2}\big) \, \theta(T-R)\, .
\end{equation}
Finally, combining (\ref{eq:Gintegral1}) with (\ref{eq:Irtbeta}),
and taking into account that for $R>0$, 
\begin{equation}
\theta(T)\theta(T-R)=\theta(T-R),
\quad
\theta(T)\delta(T-R)=\delta(T-R), 
\end{equation}
we get the Green's function,
\begin{equation}
\label{eq:GRTfinal}
G(R,T)
=
\left(\frac{\delta(T-R)}{4\pi R} 
+ 
\frac{ \beta}{4\pi}
\frac{\theta(T-R) \, I_1\big(\beta \sqrt{T^2 - R^2}\big)  }{\sqrt{T^2 - R^2}} \right)e^{-{\beta}T}\, .
\end{equation}

To recover the $\beta \to 0$ limit, we first recall that the modified 
Bessel function $I_1(z)$ has the expansion for small $z$,
\begin{equation}
I_1(z) = \frac{z}{2} + O(z^3), \quad z \to 0.
\end{equation}
Here, the argument of the Bessel function is
\[
z = \beta \sqrt{T^2 - R^2}.
\]
Thus, for small $\beta \ll 1$,
\begin{equation}
I_1\Big(\beta \sqrt{T^2 - R^2}\Big) \approx \frac{\beta \sqrt{T^2 - R^2}}{2},
\end{equation}
and the second term of $I(r,t;\beta)$ becomes
\begin{align}
\frac{\pi \beta R}{2 \sqrt{T^2 - R^2}} \, I_1\Big(\beta \sqrt{T^2 - R^2}\Big) \, \theta(T-R)
&\approx \frac{\pi \beta R}{2 \sqrt{T^2 - R^2}} \cdot \frac{\beta \sqrt{T^2 - R^2}}{2} \, \theta(T-R) 
= \frac{\pi \beta^2 R}{4} \, \theta(T-R).
\end{align}
In the limit $\beta \to 0$ we get
\begin{equation}
\lim_{\beta \to 0} \frac{\pi \beta R}{2 \sqrt{T^2 - R^2}} \, I_1\Big(\beta \sqrt{T^2 - R^2}\Big) \, \theta(T-R) = 0,
\end{equation}
and the integral reduces to a simple delta function,
\begin{equation}
\lim_{\beta \to 0} I(R,T;\beta) = \frac{\pi}{2} \, \delta(T-R).
\end{equation}

Our derivation shows that the familiar delta function contribution 
arises naturally from the derivative structure of the integral 
$I(R,T;\beta)$ and survives the $\beta \to 0$ limit, whereas the 
Bessel-function contribution, which describes the effect of finite 
damping, vanishes. Consequently, the Green's function in the 
``dampless'' case exhibits the familiar sharp propagation along 
the light cone $T = R$.

\subsection{Remarks}

1. It is useful to compare our derivation with the corresponding 
calculation presented in the Appendix to Ref.\ \cite{Tautz2016} 
[open access at:  \url{https://iopscience.iop.org/article/10.1088/1674-4527/16/10/162} ], 
in which the telegraph equation and its solution appear in a different context 
(see, particularly, Eqs.\ (A.1), (A.3), and (A.25)). In the final expression for 
the Green's function, Eq.\ (A.25), the Heaviside factor $\theta(t-r)$ and, 
most importantly, Dirac's $\delta(t-r)$ term seem to be missing. 
In contrast, in our formula, Eq.\ (\ref{eq:GRTfinal}), both contributions 
are present, which help recover the correct $\beta \to 0$ limit.

2. At the time of writing, the Wikipedia article \url{https://en.wikipedia.org/wiki/Green's_function} 
shows Green’s function for the relativistic heat conduction equation 
(see the last row in the summary table there) which is different from 
our Eq.\ (\ref{eq:GRTfinal}). 
A related discussion, \url{https://mathoverflow.net/questions/343438/greens-function-for-3d-relativistic-heat-equation}, 
points out a problem with the Wikipedia's formula and cites \cite{Tautz2016}. 
It is hoped that the derivation presented here may help shed new light on that 
conversation.

\section{Mathematical Origin of the Discrepancy}

To clarify the discrepancy highlighted in Remark 1, we explicitly examine 
the step in Ref.~\cite{Tautz2016} where the error was introduced. 
The authors evaluated an integral equivalent to our $I(R,T;\beta)$ by 
taking the derivative of a known integral with respect to a spatial parameter.

Specifically, they utilized the relation,
\begin{equation}
I(R,T) \sim -\frac{\partial}{\partial b} F(a,b,c)
\end{equation}
where $F(a,b,c)$ (in their notation) corresponds to the integrated Bessel 
form $\frac{\pi}{4} I_0(\dots)$. However, in performing this differentiation, 
the authors differentiated the continuous Bessel function part of the 
expression but neglected the derivative of the Heaviside step function 
$\theta(a-b)$ that implicitly limits the domain. 

Mathematically, if $F(b) = f(b)\theta(a-b)$, the derivative is
\begin{equation}
\frac{\partial F}{\partial b} = f'(b)\theta(a-b) - f(b)\delta(a-b).
\end{equation}
The first term corresponds to the ``wake'' (the Bessel $I_1$ term) found 
in Ref.~\cite{Tautz2016}. The second term corresponds to the wavefront 
(the Dirac $\delta$ term). By neglecting the second term, the derived 
Green's function in \cite{Tautz2016} loses the ballistic component required 
for causality and the correct undamped limit.

In addition to the missing singular wavefront discussed above, the smooth 
``wake'' component of the Green's function in Ref.~\cite{Tautz2016} 
contains a distinct algebraic error. Even if the Heaviside step function 
were restored to Eq.~(A.25) of \cite{Tautz2016}, the remaining functional 
form is physically incorrect.

This error can be identified by tracing the radial variables. In the derivation, 
the Green's function arises from a spherical integral of the form:
\begin{equation}
G \sim \frac{1}{R} \frac{\partial}{\partial R} \left[ I_0\left(\beta\sqrt{T^2-R^2}\right) \right].
\end{equation}
The derivative of the Bessel function introduces a factor proportional 
to the inner derivative of the argument:
\begin{equation}
\frac{\partial}{\partial R} \sqrt{T^2-R^2} = \frac{-R}{\sqrt{T^2-R^2}}.
\end{equation}
Consequently, the factor of $R$ in the numerator of the derivative cancels 
exactly with the factor of $1/R$ in the prefactor of the Green's function. 
The correct coefficient for the wake term is therefore independent of $R$ 
(and $T$) in the numerator, scaling simply as $\beta$.

In contrast, Eq.~(A.25) of Ref.~\cite{Tautz2016} contains a prefactor of 
$(t-t')$ in the numerator:
\begin{equation}
G_{\text{TL}} \propto \frac{t-t'}{\sqrt{(t-t')^2 - \varrho^2}} I_1(\dots).
\end{equation}
This linear time dependence is algebraically unjustified and leads to 
incorrect asymptotic behavior. 

We can verify this by considering the long-time limit ($t \to \infty$), 
where the telegraph equation must relax to the diffusion equation. 
The Green's function for three-dimensional diffusion is known to decay 
as $t^{-3/2}$. Using the asymptotic expansion of the modified Bessel 
function, $I_1(z) \approx e^z / \sqrt{2\pi z}$ for large $z$, we verify 
the behavior of the corrected solution Eq.~(\ref{eq:GRTfinal}),
\begin{equation}
G_{\text{correct}} \sim e^{-\beta t} \frac{1}{t} \frac{e^{\beta t}}{\sqrt{t}} \sim t^{-3/2},
\end{equation}
which correctly recovers the diffusive limit.
However, the solution in Ref.~\cite{Tautz2016}, due to the extra factor of $t$, 
scales as
\begin{equation}
G_{\text{TL}} \sim e^{-\beta t} \frac{t}{t} \frac{e^{\beta t}}{\sqrt{t}} \sim t^{-1/2}.
\end{equation}
This slow decay ($t^{-1/2}$) violates particle conservation laws for a 
diffusive process in three dimensions, further confirming that the 
``wake'' term in the published result is incorrect.

\end{document}